\RequirePackage{fix-cm}

\documentclass[twocolumn,final]{svjour3}       

\usepackage[dvipdfmx]{graphicx}

\usepackage{times} 
\usepackage{mathptmx}      
\usepackage{amssymb}

\usepackage[utf8]{inputenc} 
\usepackage[T1]{fontenc} 

\journalname{Applied Physics A}

\begin{document}

\title{Gas-atomized particles of giant magnetocaloric compound HoB$_{2}$
for magnetic hydrogen liquefiers}

\titlerunning{Gas-atomized particles of giant magnetocaloric compound HoB$_{2}$}

\author{Takafumi D. Yamamoto \and Hiroyuki Takeya \and Akiko T. Saito \and Kensei Terashima
\and Pedro Baptista de Castro \and Takenori Numazawa \and Yoshihiko Takano}

\institute{
Takafumi D. Yamamoto \and Hiroyuki Takeya \and Akiko T. Saito \and Kensei Terashima
\and Pedro Baptista de Castro \and Takenori Numazawa \and Yoshihiko Takano
		\at National Institute for Materials Science, Tsukuba, Ibaraki 305-0047, Japan\\
		\email{YAMAMOTO.Takafumi@nims.go.jp}	
		\and
	    Pedro Baptista de Castro \and 	    Yoshihiko Takano
		\at University of Tsukuba, Tsukuba, Ibaraki 305-8577, Japan
}

\date{Received: date / Accepted: date}

\maketitle

\begin{abstract}
The processing of promising magnetocaloric materials into spheres is
one of the important issues on developing high-performance magnetic refrigeration systems.
In the present study, we achieved in producing spherical particles of
a giant magnetocaloric compound HoB$_{2}$
by a crucible-free gas atomization process,
despite its high melting point of 2350 $^{\circ}$C.
The particle size distribution ranges from 100 to 710 $\mu$m centered at 212-355 $\mu$m
with the highest yield of 14-20wt\% of total melted electrode,
which is suitable for magnetic refrigeration systems.
The majority of the resulting particles are mostly spherical
with no contamination during the processing,
while unique microstructures are observed on the surface and inside.
These spherical particles exhibit sharp magnetic transitions
and huge magnetic entropy change of 0.34 J cm$^{-3}$ K$^{-1}$
for a magnetic field change of 5 T at 15.5 K.
The high sphericality and the high magnetocaloric performance suggest that
the HoB$_{2}$ gas-atomized particles have good potential
as magnetic refrigerants for use in magnetic refrigerators for hydrogen liquefaction.

\keywords{Magnetocaloric effect \and Gas-atomization \and Rare earth compounds \and hydrogen liquefaction}

\end{abstract}

\section{Introduction}
Magnetic refrigeration is a promising cooling technology
for replacing conventional vapor compression refrigeration.
It is based on the magnetocaloric effect (MCE) of magnetic materials
and has the advantages of high efficiency, energy saving, and environmentally friendliness
\cite{Zimm-ACE-1998,Tegus-Nature-2002,Bruck-JPD-2005,Gschneidner-RPP-2005,Lyubina-JPD-2017,Franco-PMS-2018}.
Among potential applications, there is an increasing interest
in hydrogen liquefaction by magnetic refrigeration
\cite{Numazawa-Cryo-2014},
because liquid hydrogen is one of the efficient form for transportation and storage
in the so-called hydrogen society in which hydrogen is used as an energy carrier
\cite{Sherif-IJHE-1997}.
In order to establish the magnetic refrigeration technology for hydrogen liquefaction,
much effort has been devoted to the search for new magnetic materials with a large MCE
\cite{Zhang-PhysicaB-2019,Li-SSC-2014,Matsumoto-JMMM-2017,Omote-Cryogenics-2019}
and the development of highly-efficient refrigeration systems
using an active magnetic regenerators (AMR) cycle
\cite{Kamiya-Cryoc-2007,Utaki-Cryoc-2007,Matsumoto-JPCS-2009,Kim-Cryogenics-2013}.

On the other hand, the candidate materials should be processed into spheres
for practical use in the AMR system
\cite{Yu-IJR-2010,Nielsen-IJR-2011,Tusek-IJR-2013}.
A large surface area of spheres is desirable for getting a good heat exchange
between spheres and a heat-exchanger fluid,
which is important for better performance of an AMR system.
The smaller particle size, the higher heat exchange efficiency,
but the concomitant increase in pressure loss leads to poor performance.
Considering the trade-off between the two,
it is derirable that the particle diameter is on the order of submilimeter \cite{Barclay-CPE-1984}. 

In 2020, the feromagnetic material HoB$_{2}$ has been found to exibit a giant MCE
near the liquefaction temperature of hydrogen (20.3 K) \cite{Pedro-NPG-2020};
The magnetic entropy change ($\Delta S_{M}$) at a Curie temperature ($T_{\rm C}$) of 15 K has 
a maximum value of 40.1 J kg$^{-1}$ K$^{-1}$ (0.35 J cm$^{-3}$ K$^{-1}$)
for a magnetic field change of 5 T,
which is greater than those observed in any bulk materials
that have been studied for magnetic hydrogen liquefiers.
Due to such the $\Delta S_{M}$ and $T_{\rm c}$,
HoB$_{2}$ is expected to work as suitable magnetic refrigerants for hydrogen liquefaction.
However, the material properties of this boride make it difficult
to process it into the desired spherical shape.
For instance, HoB$_{2}$ has a high melting point of 2350 $^{\circ}$C
comparable to those of ceramics.
Accordingly, conventional atomization techniques \cite{Antony-JOM-2003}
that melt and/or superheat a target material in a ceramic crucible cannot be used.
Besides, the plasma rotating electrode process \cite{Miller-Cryoc-2001,Fujita-JJAP-2007} is
a non-contact type method to produce a high-purity spherical particles,
but the brittle nature of HoB$_{2}$ hinders making an electrode
and using it with high speed rotations during the processing.

In this study, we succeed in fabricating HoB$_{2}$ spherical particles for the first time
by using an electrode induction melting gas atomization (EIGA) technique,
in which the electrode rod of a material is melted without the use of the crucible.
The resulting particles are evaluated in terms of morphology, phase confirmation,
and various physical properties including magnetocaloric properties.

%
%
\section{Experimental details}
\begin{figure}[b]
\centering
\includegraphics[width=85mm,clip]{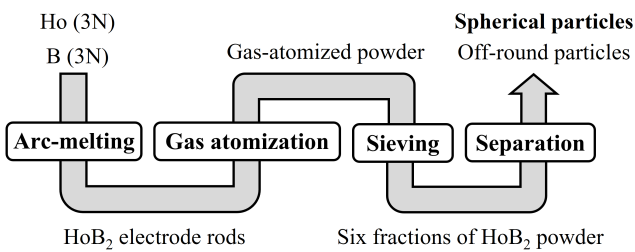}
\caption{(Color Online) The diagram of preparation process of HoB$_{2}$ spherical particles.}
\label{fig:Diagram}
\end{figure}
\begin{figure}[t]
\centering
\includegraphics[width=75mm,clip]{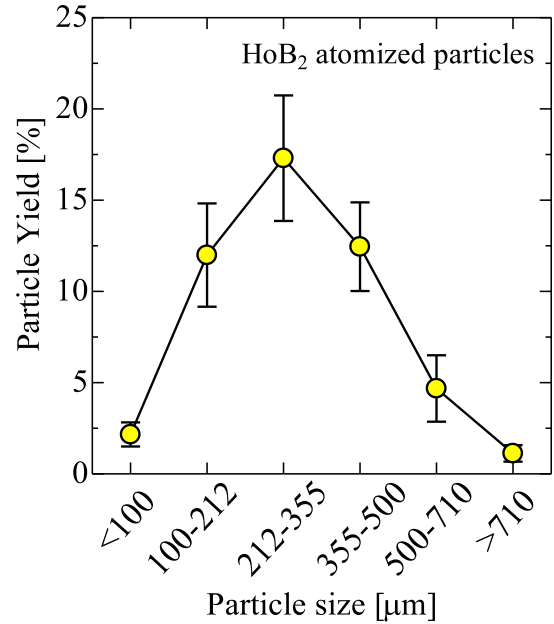}
\caption{(Color Online) Particle size distribution of HoB$_{2}$ spherical particles.}
\label{fig:EIGA}
\end{figure}

Figure \ref{fig:Diagram} depicts the diagram of preparation process of HoB$_{2}$ spherical particles.
The HoB$_{2}$ electrode rods were made by arc-melting
in a water-cooled copper heath arc furnace.
Stoichiometric amounts of Ho (3N) and B (3N) elements were arc-melted under an Ar-atmosphere.
Each rod was flipped and melted several times for homoginizing the sample.
The resulting electrode rods were 100-300 g in weight.
In the present EIGA process, one end of HoB$_{2}$ electrode rod was
fixed with a chuck in an Ar-atmosphere.
The rod was immersed into an induction coil and inductively melted at the other end.
The molten metal then freely fell through the orifice,
atomized by jetting Ar-gas, and solidified into spheres.
After the atomization, the collected powder was sieved into six fractions,
$<$100 $\mu$m, 100-212 $\mu$m, 212-355 $\mu$m, 355-500 $\mu$m, 500-710 $\mu$m, and $>$710 $\mu$m,
through a series of JIS Z 8801 standard sieves with a FRITSCH vibratory sieve shaker ANALYSETTE 3.
The sieved powder consisted of spherical particles and off-round ones,
so the former were separated from the latter by rolling them on a sloped belt conveyor.

Microstructural observation was carried out
using a Hitachi SU-70 scanning electron microscope (SEM) operated at 20 kV.
Powder X-ray diffraction (XRD) measurements were performed at rooom temperature
by a Rigaku MiniFlex600 diffractometer with Cu K$\alpha$ radiation.
The temperature dependence of magnetization ($M$-$T$ curves)
at various magnetic fields ($\mu_{0} H$) ranging from 0.01 to 5 T were
measured between 2 and 50 K in field cooling processes
by a Quantum Design SQUID magnetometer.
The isothermal magnetization curves ($M$-$H$ curves)
were collected at 2 K between 0 and 5 T.
The magnetic entropy change was evaluated from a series of the $M$-$T$ curves
according to a Maxwell's relation
\begin{equation}
\Delta S_{M}(T, \mu_{0}\Delta H) =
\mu_{0} \int_{0}^{H}{\left(\frac{\partial M}{\partial T}\right)_{H}}dH,
\end{equation}
where $\mu_{0}\Delta H$ is the magnetic field change from zero to $\mu_{0} H$.
In the magnetization measurements, 
the bulk sample was formed into a rectangular shape
with dimensions of 2.4$\times$0.5$\times$0.5 mm${^3}$,
and dozens of the spherical particles in contact with each other were
arranged vertically long with the aspect ratio of about 2-4.
The magnetic field was applied along the longitudinal directions of
both the rectangular and the arranged spheres.
The specific heat data at 0 T were measured by a thermal relaxation method
with a Quantum Design PPMS.
%
%
\section{Results and discussion}
Figure \ref{fig:EIGA} shows the particle size distribution for HoB$_{2}$ spherical particles.
The gas atomization process was performed repeatedly,
and the average values of the resulting yield are shown here.
We find that the yield of the spherical particles has a significant value
in the range from 100 to 710 $\mu$m,
where the distribution patterns are centered around 212-355 $\mu$m
with the yield of 14-20wt\% of the total melted electrode.
The obtained particle size range is suitable for AMR systems.
This may be due to the fact that
the atomizing gas pressure in this study was set to 1.5-3.5 MPa,
which is lower than the values used in conventional EIGA experiments for producing fine powder
\cite{Franz-Titanium-2008,Xie-IOPconf-2019,Sun-IOPconf-2019},
in which the particle size is typically 100 $\mu$m or less
\cite{Franz-Titanium-2008,Xie-IOPconf-2019,Sun-IOPconf-2019,Feng-ChiPhysB-2017,Guo-ActaMetall-2017}.

\begin{figure}[t]
\centering
\includegraphics[width=85mm,clip]{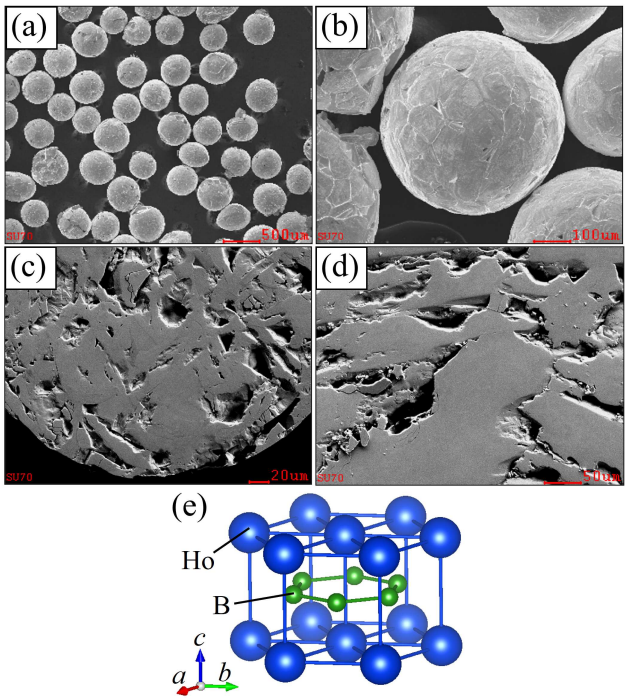}
\caption{(Color Online) (a) SEM image of surface and (b) its enlarged image
for HoB$_{2}$ particles with 355-500 $\mu$m diameter.
The cross-sectional SEM image of (c) the HoB$_{2}$ particles and (d) the HoB$_{2}$ electrode.
(e) The crystal structure of HoB$_{2}$ with the space group of \textit{P}6/\textit{mmm}.
}
\label{fig:SEM}
\end{figure}

Figure \ref{fig:SEM}(a) shows a SEM image of surface
for HoB$_{2}$ particles with 355-500 $\mu$m diameter.
The majority of the particles are mostly spherical
and has no satellites as is seen in other gas-atomized materials
\cite{Guo-ActaMetall-2017,Mayer-PSS-2014,Osborne-ACE-1994}.
By using an image analysis software Image-J (National Institute of Health, US),
the roundness of these particles is evaluated to be about 0.88.
This quantity, defined as $4\times (\rm {Area})/(\pi \times(\rm {Major \ axis})^2)$,
equals 1 for a circular object and less than 1 for an object that departs from circularity.
The obtained value of the roundness gurantees
the good sphericality of HoB$_{2}$ gas-atomized particles.

As shown in Fig. \ref{fig:SEM}(b), however,
the surface of each particle is not smooth but has a turtle shell-like structure.
Moreover, a more complex inner structure with voids can be seen in Fig. \ref{fig:SEM}(c).
It should be noted that such internal voids do not result from the gas atomization process,
because similar internal texture is observed in the HoB$_{2}$ electrode (Fig. \ref{fig:SEM}(d)).
These microstructures on the surface and inside can be attributed to the crystal structure of HoB$_{2}$.
This diboride has a layered structure with a hexagonal B-network
as shown in Fig. \ref{fig:SEM}(e),
so the crystals are expected to grow while being oriented along the \textit{ab}-plane.
When the sample is solidified relatively quickly
in the gas-atomization process and the arc-melting process,
it is possible that the crystals are oriented inhomogeneously
and consequently the microstructures are formed.

\begin{figure}[t]
\centering
\includegraphics[width=85mm,clip]{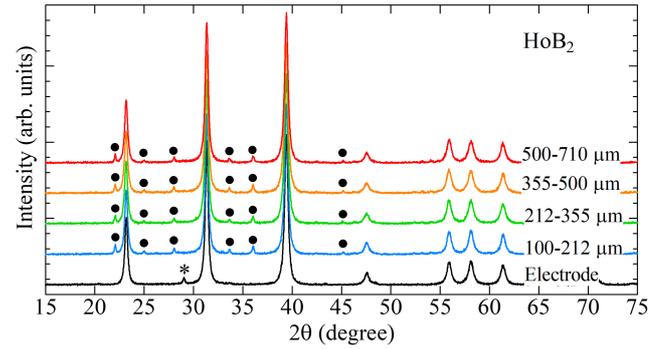}
\caption{(Color Online) Powder X-ray diffraction patterns of HoB$_{2}$ electrode
and HoB$_{2}$ particles with various diameters from 100 to 710 $\mu$m.
The marks indicate the Bragg peaks of Ho$_{2}$O$_{3}$ ($\ast$) and HoB$_{4}$ ($\bullet$).}
\label{fig:XRD}
\end{figure}

The powder XRD patterns shown in Fig. \ref{fig:XRD} confirm that
both HoB$_{2}$ spherical particles and HoB$_{2}$ electrode have
the main phase of the hexagonal HoB$_{2}$
with the space group of \textit{P}6/\textit{mmm}.
The position and width of HoB$_{2}$ peaks are the same before and after atomizing,
which suggests no change in crystallographic properties of HoB$_{2}$ phase
by the atomization process.
Furthermore, no differences in the peaks are found in all the atomized samples.
Even though the solidification rate should depend on the particle diameter,
it does not seem to affect the crystal structure of HoB$_{2}$.
As an impurity phase, a tiny amount of Ho$_{2}$O$_{3}$ is found for the electrode
while it disappears after atomizing.
Instead, a bit trace of of HoB$_{4}$ is observed in all the atomized particles.
HoB$_{2}$ is a peritectic system
in which it decomposes into solid HoB$_{4}$ and a Ho-rich liquid phase at 2200 $^{\circ}$C
before completely melted at 2350 $^{\circ}$C \cite{Liao-PD-1990}.
Thus, HoB$_{4}$ probably appeared
when the HoB$_{2}$ electrode was melted in the atomization process.
Nevetheless, there are no other impurities,
implying that it is free from contamination during the processing.

\begin{figure}[t]
\centering
\includegraphics[width=85mm,clip]{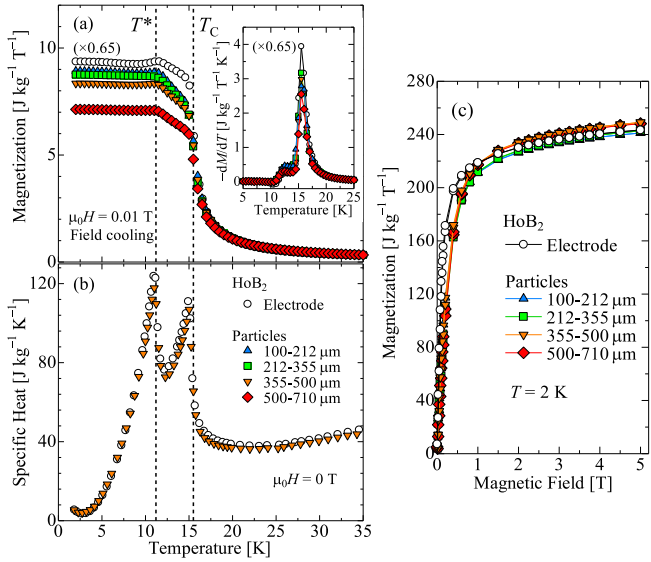}
\caption{(Color Online)
Temperature dependence of (a) magnetization at 0.01 T
and (b) specific heat at 0 T in HoB$_{2}$ electrode and HoB$_{2}$ particles.
(c) Isothermal magnetization curves at 2 K.
The inset in (a) depicts the temperature derivative of magnetization d$M$/d$T$ at 0.01 T.}
\label{fig:Properties}
\end{figure}

Now let us investigate the physical properties of the HoB$_{2}$ atomized particles.
Figure \ref{fig:Properties}(a) shows the $M$-$T$ curves at 0.01 T in field cooling processes.
Here the data for the electrode is scaled by 0.65 for visibility.
In both the electrode and all the particles with different diameters,
the magnetization rapidly increases as the temperature decreases toward 15 K,
which is indicative of a ferromagnetic transition.
The Curie temperature $T_{\rm C}$, defined as a peak temperature of d$M$/d$T$, is
evaluated to be 15.5 K for all the samples (see the inset of Fig. \ref{fig:Properties}(a)).
Furthermore, all the $M$-$T$ curves exhibit a kink anomaly around 11 K,
which is associated with the spin-reorientation phenomenon \cite{Terada-PRB-2020}.
The kink temperature, named by $T^{\ast}$, does not change
before and after the atomization process.
As shown in Fig. \ref{fig:Properties}(b),
the well-defined peaks in specific heat at 0 T are observed
around $T_{\rm C}$ and $T^{\ast}$,
meaning that clear phase transitions take place at each temperature.
These peaks in the HoB$_{2}$ particles are as sharp as those in the electrode.
In addition, the $M$-$H$ curves at 2 K for all the samples have
a simialr magnetic field dependence
and quantitatively agree with eath other in high magnetic fields
(Fig. \ref{fig:Properties}(c)).
All these results suggest that the HoB$_{2}$ spherical particles are homogeneous
and of good quality comparable to the bulk counterpart.

The disagreement in $M$-$H$ curves below 1 T between the electrode and the particles should
reflect the extrinsic demagnetization effect:
The lower aspect ratio of the arranged particles can result
in a larger demagnetization field.
The difference between the $M$-$T$ curves is also probably due to that of the aspect ratio.
Accordingly, we conclude that there is no intrinsic dependence of
magnetic properties on the particle diameter.
In the gas-atomization process, the quenching effect on atomized particles is expected,
and the cooling time of atomized particles can vary by diameter.
No diameter dependence implies that the quenching effect has no impacts on
the magnetic properties of HoB$_{2}$.
This is consistent with no differences in physical properties
between the electrode and the particles.
Note that it is difficult at this stage
to discuss the relationship of the quenching effect
with the physical propreties in depth,
because the nature of HoB$_{2}$ itself has not been clarified.

It should be also discussed whether the impurity HoB$_{4}$ affects
the physical properties of the HoB$_{2}$ particles.
This tetraboride is known to exhibit two antiferromagnetic-like transitions at 7.1 K and 5.7 K,
at which the magnetization shows a kink or drop and the specific heat shows sharp peaks
\cite{Fisk-SSC-1981,Kim-JAP-2009}.
As is evident in Figs. \ref{fig:Properties} (a) and \ref{fig:Properties} (b),
no such anomalies are found in the data for the HoB$_{2}$ particles.
Thus, it is likely that HoB$_{4}$ with a small amount has no influence
on the final physical properties of HoB$_{2}$.
This is true for a small amount of Ho$_{2}$O$_{3}$,
which has an antiferromagnetic transition temperature of 2 K \cite{Boutahar-SciRep-2017}.

\begin{figure}[t]
\centering
\includegraphics[width=85mm,clip]{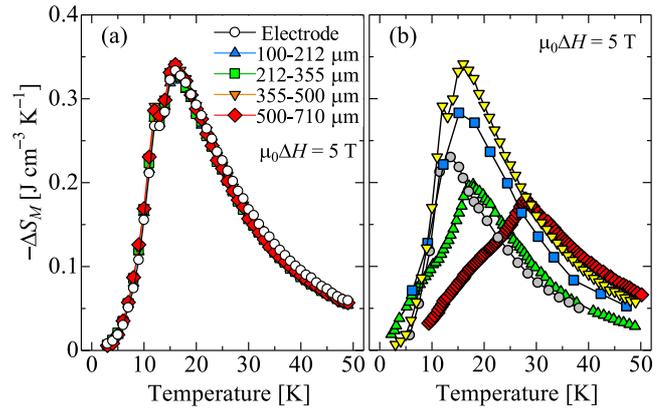}
\caption{(Color Online) (a) Magnetic entropy change for $\mu_{0} \Delta H =$ 5 T
in HoB$_{2}$ particles and HoB$_{2}$ electrode.
(b) The comparison of the HoB$_{2}$ particles with 355-500 $\mu$m diameter ($\bigtriangledown$)
and other promising magnetocaloric materials:
ErAl$_{2}$ ($\bigcirc$) \cite{Pecharsky-JAP-1999},
HoN ($\square$) \cite{Yamamoto-JAC-2004},
EuS ($\bigtriangleup$) \cite{Matsumoto-Cryogenics-2016},
and HoAl$_{2}$ ($\Diamond$) \cite{Hashimoto-JAP-1987}.}
\label{fig:DeltaS}
\end{figure}

Finally, we will evaluate the magnetocaloric properties of the HoB$_{2}$ atomized particles.
Figure \ref{fig:DeltaS}(a) shows the magnetic entropy change $\Delta S_{M}$
for $\mu_{0}\Delta H =$ 5 T in the spherical particles and the electrode.
Here we take J cm$^{-3}$ K$^{-1}$ as the units of $\Delta S_{M}$
as this unit is meaningful from an engineering point of view \cite{Gschneidner-RPP-2005}.
Besides, it has shown that the demagnetization effect does not seriously affect
the evaluation of $\Delta S_{M}$ for the large $\mu_{0}\Delta H$ \cite{Bez-JMMM-2018}.
The $\Delta S_{M}$ is qualitatively and quantitatively the same
between all the spherical particles and the bulk counterpart:
It peaks at $T_{\rm c}$ and $T^{\ast}$
and takes the maximum value of about 0.34 J cm$^{-3}$ K$^{-1}$ at $T_{\rm c}$.
These results are almost consistent with those in our previous work \cite{Pedro-NPG-2020}.
Therefore, it is found that HoB$_{2}$ spherical particles prepared in this study have
the same magnetocaloric properties as bulk material.
Furthermore, we would like to compare here HoB$_{2}$
with other promising magnetocaloric materials
that have similar magnetic transition temperatures.
It can be seen from Fig. \ref{fig:DeltaS}(b) that
the $\Delta S_{M}$ of HoB$_{2}$ is not only large in peak value,
but also has the advantage of keeping a large value up to high temperatures.
It is worth noting that the values of $\Delta S_{M}$ in HoB$_{2}$ are 
comparable to those in HoAl$_{2}$ even at above 30 K.
This feature of HoB$_{2}$ is attractive as magnetic refrigerants
in that it is able to work over a wide temperature range.
%
%
\section{Conclusion}
We have suceeded in producing the spherical particles of
a giant magnetocaloric compound HoB$_{2}$
by the electrode induction melting gas atomization technique.
The obtained particle size range is suitable for AMR systems:
it ranges from 100 to 710 $\mu$m centered at 212-355 $\mu$m
with the highest yield of 15-25wt\% of total melted electrode. 
The resulting particles have almost perfect spherical shape
and no contamination during the processing.
Furthermore, they have the characteristic microstructures on the surface and inside,
which may originate from the crystallographic nature of HoB$_{2}$.
The physical properties of the spheres are
quite similar to those of the bulk counterpart,
in which the sharp magnetic transitions
and the giant magnetic entropy change are observed.
All the results suggest that
the gas-atomized HoB$_{2}$ particles have good potential as magnetic refrigerants
for use in magnetic hydrogen liquefiers.

Not limited to HoB$_{2}$, magnetocaloric materials are often high melting point materials,
which may hinder processing them into particles by conventional atomization methods.
Our experimental results have demonstrated that
the present atomization process is a viable route to fabricate spherical particles,
even for the materials with melting points above 2000 $^{\circ}$C.
This fact would encourage the production of particles for various magnetocaloric compounds,
which makes a significant contribution for developing magnetic refrigeration systems.

\begin{acknowledgements}
This work was supported by JST-Mirai Program Grant Number JPMJMI18A3, Japan.
\end{acknowledgements}

\section*{Complince with ethical standards}
\subsection*{Conflict of interest}
The authors declare that they have no conflict of interest.


\end{document}